\documentclass[eps,12pt]{article}
\usepackage{graphicx}

\begin{document}

% SETS A DOUBLE SPACE
\baselineskip=18pt

\begin{center}
{\Large\bf Partition function zeros and leading order 
       scaling correction of the 
       3D Ising model from multicanonical simulations}
\vskip 1.1cm
{\bf Nelson A. Alves\dag \footnote{E-mail: alves@quark.ffclrp.usp.br}
           \, J. R. Drugowich de Felicio\dag \footnote{E-mail:  
                                    drugo@pinguim.ffclrp.usp.br} \\
           \, and \, 
             Ulrich H.E. Hansmann\ddag \footnote{E-mail: hansmann@mtu.edu}}
\vskip 0.1cm
\dag{\it Departamento de F\'{\i}sica e Matem\'atica, FFCLRP 
     Universidade de S\~ao Paulo. Av. Bandeirantes 3900. 
      CEP 014040-901 \, Ribeir\~ao Preto, SP, Brazil}
\vskip 0.1cm
\ddag{\it Department of Physics, Michigan Technological University,
         Houghton, MI 49931-1291, USA}

\vskip 0.3cm
\today
\vskip 0.4cm
\end{center}
%%%%%%%%%%%%%%%%%%%%%%%%%%%%%%%%%%%%%%%%%%%%%%%%%%%%%%%%%%%%%%%
\begin{abstract}
   The  density of states for the 
three-dimensional Ising model is calculated with high-precision
from multicanonical simulations.
   This allows us to estimate the leading
partition function zeros for lattice sizes up to $L=32$. 
   Combining previous statistics for smaller lattice sizes, we have
evaluated the correction to scaling and the critical exponent 
$\nu$ through out an analysis of a multi-parameter fit
and of Bulirsch-Stoer (BST) extrapolation algorithm.
   The performance of the BST algorithm is also explored in case 
of the 2D Ising model, where the exact partition function zeros
are known.
\end{abstract}
\vskip 0.1cm
{\it Keywords:} Ising model, scaling exponent, 
                multicanonical simulation, partition function zeros  
\vskip 0.1cm
{\it PACS-No.: 05.50.+q, 05.70.-a, 64.60.Fr }

%%%%%%%%%%%%%%%%%%%%%%%%%%%%%%%%%%%%%%%%%%%%%%%%%%%%%%%%%%%%%%%%%
%\begin{multicols}{2}
%\twocolumn
\newpage

\section{Introduction} 
\vspace*{-0.5pt}
\noindent

In recent years  there has been a persistent interest in obtaining
accurate estimates of critical parameters of the three-dimensional 
(3D) Ising-like systems through high performance simulations 
and perturbative expansions
\cite{gupta1,blote2,blote1,gupta2,hasenbusch3,hasenbusch2,zinn,parisi,hasenbusch1,campos}.
 Our aim here is to
enlarge  the knowledge about critical behavior of Ising-like 3D systems
by calculating not only the critical exponent of the correlation
length $\nu $, but also by tackling the much harder
problem of calculating the first correction to scaling $w$.  

A common way to extract information on phase transitions from
Monte Carlo simulations is by means of finite-size scaling (FSS),
for instance by analyzing the partition function 
zeros in the complex temperature plane\cite{itzykson,r13}. 
   This approach is not restricted to 
Ising-like systems and was recently even used to study structural
transitions in bio-molecules \cite{ah99b}. 
  However, separation of
universality classes can be tricky for 3D systems, and requires 
high-precision data. 
  An important tool for obtaining such data 
was provided by Ferrenberg and Swendsen 
\cite{FS}\  who revived reweighting techniques introduced by Salsburg 
{\it et al.} \cite{salsburg}\ forty years ago. 
  In the sequence, 
 multiple-histogram \cite{swendsen,alves1}
and multicanonical simulations \cite{Mu}\ were proposed 
for a reliable  numerical determination of the density of states  
and extensively checked 
in 
two-dimensions 
where exact results are available.  
  Only by combining one of  these 
sophisticated new simulation techniques
(exhaustive multicanonical Monte Carlo simulations for $L\leq 32$) 
with Itzykson's finite-size scaling relation for the partition
function zeros \cite{itzykson}, we were able to obtain the results
presented in this article.

Before proceeding further, we give the outline of the paper. In the next
section we describe the numerical evaluation of the partition function.
  Section 3
is concerned with the crucial task of data analysis.
  There we calculate the critical exponent 
$\nu$ and the correction to scaling.
  The exponent $\nu $ was first obtained by a 
four-parameter fitting. Next, we used the Bulirsch-Stoer algorithm 
\cite{BST,s5}
to extrapolate results from finite lattices to the thermodynamical limit. 
That algorithm has a free-parameter $w$ which  is  
exactly
the 
wanted correction to scaling. We further corroborate on results recently 
obtained by other techniques. 
Finally, we  test the performance of our approach, and specially the
BST algorithm, for the 2D Ising model where  the exact zeros are known.

%%%%%%%%%%%%%%%%%%%%%%%%%%%%%%%%%%%%%%%%%%%%%%%%%%%%%%%%%%%%%%%%%
\section{Multicanonical simulation and partition function zeros}
\noindent
In the multicanonical algorithm \cite{Mu}
conformations with energy $E$ are assigned a weight
$  w_{mu} (E)\propto 1/n(E)$. Here, $n(E)$ is the density of states.
A  simulation with this weight
will  lead to a uniform distribution of energy:
\begin{equation}
  P_{mu}(E) \,  \propto \,  \rho(E)~w_{mu}(E) = {\rm const}~.
\label{eqmu}
\end{equation}
This is because the simulation generates a 1D random walk in the energy,
allowing itself to escape from any  local minimum.
Since a large range of energies are sampled, one can
use the reweighting techniques \cite{FS} to  calculate thermodynamic
quantities over a wide range of temperatures by
\begin{equation}
<{\cal{A}}>_T ~=~ \frac{\displaystyle{\int dx~{\cal{A}}(x)~w^{-1}(E(x))~
                 e^{-\beta E(x)}}}
              {\displaystyle{\int dx~w^{-1}(E(x))~e^{-\beta E(x)}}}~,
\label{eqrw}
\end{equation}
where $x$ stands for configurations.

It follows from equation~\ref{eqmu} that the multicanonical algorithm allows us
to calculate estimates for the spectral density:
\begin{equation}
  \rho(E) = P_{mu} (E) w^{-1}_{mu} (E)~.
\end{equation}
We can therefore construct
the partition function of the three-dimensional Ising model 
\begin{equation}
 Z(\beta) =  \sum_{E} \rho(E) u^E ~,              \label{eq:z1}
\end{equation}
where we define $u = e^{-4\beta}$.

In the present article, the estimates of the partition function rely
on averages over NRUN simulations of NSWEEP Monte Carlo updates for
3D Ising models of linear size $L$. In order to allow the system to
thermalize, additional sweeps in the
beginning were performed and discarded. Table 1 lists the respective values
for NSWEEP and NRUN.

Once, we have calculated reliable estimates for the partition function we
can calculate the zeros. In the polynomial form, equation (\ref{eq:z1})
has a large number of coefficients for large lattice sizes. 
 For this reason
we apply the method presented in \cite{alves2} to obtain the leading
complex zeros $u_1^0(L)$. 
In table 1 we collect the so calculated zeros as obtained from
 our multicanonical simulations.

%%%%%%%%%%%%%%%%%%%%%%%%%%%%%%%%%%%%%%%%%%%%%%%%%%%%%%%%%%%%%%%%%
\section{Finite size scaling analysis}
\noindent

The standard FSS approach for the zero $u_1^0(L)$ closest to
the real positive axis neglects, for sufficiently large $L$, 
corrections to scaling \cite{itzykson},
\begin{equation}
u_1^0(L) = u_c + A L^{-1/\nu}\left[ 1 + {\it O}(L^{-w})\right]\,. 
                                                      \label{eq:1fit} 
\end{equation}
Since we are interested in analysing $\nu$ in function of the correction
to scaling exponent $w$, we follow ref. \cite{alves1,r14}
and fit our data to a four parameter scaling relation,
\begin{equation}
|u_1^0(L)-u_c| = a_1 L^{-1/\nu} + a_2 L^{-a_3}\,.   \label{eq:4fit} 
\end{equation}
However, due to the small number of available lattices, such a  
multi-parameter fit is  not appropriate. 
To circumvent this difficulty we proceed including 
statistics already available 
for smaller lattice sizes. 
   In ref. \cite{alves1}, table 1, zeros for 3D Ising model
are presented for $L = 6,8,10$ and $14$, together
with data from ref. \cite{r14} for $L = 3,4,5,6,8$ and $10$.
 In addition, we decided to take into account that information and our
results in table 1, weighted with the corresponding available precision.
 The final estimates for each lattice is a linear combination of available
data with normalized weight factors  which 
 are taken as the reciprocals of the corresponding
empirical variances for each data \cite{s1}.

In figure 1 we show the corresponding fit for all lattice sizes, whose
parameters are found by monitoring the goodness of fit Q \cite{r16}.
Here we choose the recently obtained critical value 
$ u_c = 0.41204684(25)$  \cite{parisi},
%%\begin{equation}
%%  u_c = 0.41204684(25) \, ,                           \label{eq:uc}     
%%\end{equation} 
to apply the least-square method to equation \,(\ref{eq:4fit}),
although the fit is not very sensible to the precision in the 
value of critical temperature $u_c$. 
The error bar of those data is included, but it is hardly seen in
that scale. 
 The goodness of fit $(Q=0.89)$ reveals a very good agreement
with the data. We obtain $\nu=0.62853(35)$ and $a_3=4.861(84)$.
%%(It is impressive how this four-parameter equation fits all data.)

 If we discard the smallest sizes $L=3,4$ and 5, we obtain
$\nu=0.6280(15)$ ($Q=0.84$), corresponding to 
$y_t = 1/\nu = 1.5924(38)$, which is in remarkable agreement with previous
results \cite{blote2,zinn,parisi}. 
 However, this fit is less stable with
relation to the parameters in the second term.  This is because of 
the  presence
of rather large lattice sizes.

\subsection{Correction to scaling from RG transformation}

  In this section we now evaluate the correction
to scaling by a method which is similar to the ``finite size
phenomenological renormalization group (RG)'' analysis by Binder \cite{r24}
(which in turn was based on   Nightingale's finite size 
RG transformation \cite{s3}
for the correlation length $\xi_L$).
  Our   method to evaluate $w$ was previously used 
\cite{alves2} to analyse the 2D Ising model, and it is 
briefly recalled here.

In order to consider
the scaling relation for the longitudinal correlation length $\xi_L(\beta)$,
we assume that the system is of finite length scale $L$ in 
one direction and infinite in all other directions. 
The standard expression for the correlation exponent $\nu$ is 
now given by \cite{r22,r23}
\begin{equation}
 1+ \frac{1}{\nu_{L,L'}} \,  = \,
              {\rm ln}   \left( \frac{
               \rm{\raisebox{.4ex}{$\partial \xi_{L'}$}} /\partial\beta}
             { \rm{\raisebox{.4ex}{$\partial \xi_{L}$}} / \partial\beta}
               \right)_{\!\! \beta_c}  
             / \, {\rm ln} (\frac{L'}{L}) \,.    \label{eq:brezin}
\end{equation}
This expression is obtained from a linearization around the fixed
critical point $\beta_c$ for fixed scaling transformation $L \rightarrow L'$.
The scaling equation for the finite size longitudinal
 correlation length is given by 
\begin{equation}
\xi_L \,=\, L\, Y_{\xi}(\,(\beta-\beta_c)L^{1/ \nu}, hL^{y_H}, \tilde{u}L^{y_3})\,.
                                                \label{eq:night}
\end{equation}
   This differentiable equation includes corrections due to the 
leading bulk irrelevant scaling field $\tilde{u}$ with exponent $y_3 < 0$,
and a magnetic field dependence for the sake of completeness.

   From equation \,(\ref{eq:brezin}) and equation \,(\ref{eq:night}) one
obtains, for $h=0$,
\begin{equation}
 \frac{1}{\nu_{L,L'}} \, =\, \frac{1}{\nu}            \, 
  +\,  a_0 \frac{L'^{y_3} - L^{y_3}}{{\rm ln}(L'/L)}  \,
  +\,  b_0 \frac{L'^{2 y_3} - L^{2 y_3}}{{\rm ln}(L'/L)} \,
  +\, ... \,                   \label{eq:cross}
\end{equation}
where
$a_0$ and $b_0$ include derivatives like  
${\partial Y_{\xi}(y,z)} / {\partial y} |_{y=0,z=0} $.
With the introduction of the 
rescaling factor  $ s = L'/L$  in equation \,(\ref{eq:cross}), we can 
now evaluate $y_3$.

However an important  point remains to be answered: 
 how to estimate finite size dependence 
of $\nu$ on lattice sizes $L\, {\rm and}\, L'$ ?
This can be achieved from large enough
pairs of lattices $L$ and $L'$, with $L' > L$,
through the folowing expression for the partition function zeros:
\begin{equation}
\frac{1}{\nu_{L,L'}} \, = \, {\rm ln} \left( \frac{|u_1^0(L')-u_c|}
       {|u_1^0(L)-u_c|} \right) / \,{\rm ln} (\frac{L}{L'})\,. \label{eq:nu}
\end{equation}
  This equation defines our finite size estimators $\nu_{L,L'}$
from data in table 1. 
  A second estimate can be obtained with the replacement
$|u_1^0 - u_c|$ by its imaginary part Im$(u_1^0)$ in equation \,(\ref{eq:nu}).
  For large enough systems we have Re$\,(u_1^0) \sim u_c$, 
and both approaches should lead to the same result.

   In table 2 we present sequences of those two possible
estimates  $\nu_{L,sL}$ as a function
of the fixed rescaling factor $s = 2$, where
the second column stands for the results of equation \,(\ref{eq:nu})
and the third one for the replacement
         $|u_1^0(sL)-u_c|\,/ \,|u_1^0(L)-u_c|$  
      by ${\rm Im}\,u_1^0(sL)\,/ \, {\rm Im}\,u_1^0(L)$.

%%%%% (\ref{eq:nu})

   As $L$ increases, the values obtained by matching
pairs of lattices are expected to converge to a limiting value.
If we match our largest lattices $L=24$  and  32
we obtain from equation \,(\ref{eq:nu})
$\nu = 0.6260(66)$ whereas  $\nu = 0.6292(70)$ is obtained by using only
the imaginary part of the zeros.

Looking at the values for the crossings $(L,2L)$ with $(12,24)$
and $(16,32)$ in the second column of table 2, we realize that
our values for the real part of the corresponding zeros
are not precise enough to obtain an ordered sequence towards
its critical value $\nu$ as $L \rightarrow \infty$.
 On the other hand the estimate based on the imaginary part of
zeros seems to preserve this trend.

  Since equations \,(\ref{eq:cross}) and \,(\ref{eq:nu}) are valid only for
large enough lattice sizes, we have to discard the smallest 
values in table 2. 
  For this reason we prefer not to evaluate 
equation \,(\ref{eq:cross}) by a multi-parameter fit. 
  On the other hand, we note that this  relation 
%% \,(\ref{eq:cross}) 
with $L'=sL$, is an equation of type
\begin{equation}
 T(h) = T + a_1\,h^{w} + a_2\,h^{2w}\, + \cdots ,  \label{eq:bst2}
\end{equation}
where we identify,
\begin{eqnarray}
 y_3  & = & - w          \nonumber \\
 h    & = & 1/L           \nonumber \\
T(h)  & = & 1/\nu_{L,2L}  \nonumber \\  
\end{eqnarray}            
and $T= 1/\nu$ the asymptotic limit for $h \rightarrow 0$.
Hence, equation \,(\ref{eq:bst2}) is in the proper form to be analysed by
the so called BST approximants, on which we elaborate in the
next section.

\subsection{BST extrapolation}

Bulirsch and Stoer \cite{BST} developed an algorithm to extrapolate
a sequence $h_N$, $(N=0,1,2, ...)$ converging to zero as $N \rightarrow 
\infty$.  See also ref. \cite{creswick} for a recent discussion 
on BST algorithm. 

The BST algorithm approximates tabulated data $T(h)$
by a sequence of rational functions \cite{BST,s5}.
 The limiting value $T$ is computed from a table of recurrent
relations defined from:
\begin{eqnarray}
T_{-1}^{(N)} & = & 0        \nonumber  \\
T_0^{(N)} & = & T(h_N)   \label{eq:bst3}
\end{eqnarray}

and

%%\begin{equation}
\begin{eqnarray}
\lefteqn{T_m^{(N)} = T_{m-1}^{(N+1)} + (T_{m-1}^{(N+1)} - T_{m-1}^{(N)}) }
                                                              \nonumber \\
& &            \left[ \left(\frac{ h_N}{h_{N+m}}\right)^w
               \left( 1- \frac{T_{m-1}^{(N+1)} - T_{m-1}^{(N)}}
                           {T_{m-1}^{(N+1)} - T_{m-2}^{(N+1)}}
               \right) - 1 \right]^{-1}   \, .             \label{eq:bst4} 
\end{eqnarray}
%%\end{equation}
Here $w$ plays the role of a free parameter. 
If one defines $\varepsilon_m^{(i)} = 2 (T_m^{(i+1)} - T_m^{(i)}) $,
it is expected  $|T_m^{(i)} - T| \leq \varepsilon_m^{(i)}$ in the
limit $i \rightarrow \infty$. 
 The above remark gives a criterion \cite{s5}  for choosing $w$ 
as the value to minimize  $\varepsilon_m^{(i)}$,
in order to have a fast and reliable convergence.

Our aim is to extrapolate the finite size sequence in table 2 by the
BST algorithm.
 However, before proceeding with our analysis, we would like to 
explore first how the  BST extrapolant approach performs for exact data.
For this end, we come to the 2D Ising model. In ref. \cite{alves2}
the exact values for $\nu_{L,2L}$ are presented.
We show in figures 2, 3, and 4 the BST extrapolants obtained from
 different sets of sequences $T(h_N)$.
 
In figure 2 we show the BST estimates of the critical 
exponent $\nu^{\rm BST}$ from a sequence for lattices of 
lengths $L = h^{-1}$, with
 $h = 1/4,\, 1/6,\, 1/8,$ $ 1/10,\,1/12,\, 1/16,\, 1/20,\, 1/24$
and $1/32$.
 This figure presents a pole behavior at $w \sim 1.580$ for the
extrapolated results (full lines) to 
$h \rightarrow 0\, (L \rightarrow \infty)$ 
for this sequence. 
 In the neighbourhood of that pole we note the corresponding large
values for the systematic error (dashed lines) according to the 
scale at the right hand side of that figure.
 Here we define the error as the difference between the extrapolated
value $T$ and a value of last but one interaction: 
$T - T_{m-1}^{(2)}$.

 We also observe that the known value $\nu=1$ is obtained
for $w \simeq 1$ with error $\simeq 0$.
Moreover, it is remarkable that for this sequence of lengths $h$,
$\nu^{\rm BST}$ is weakly dependent on $w$.
  For instance, we obtain $\nu$ with $0.1\%$ precision, 
$\nu\, \epsilon \, [0.999,\, 1.001]$   for a large range of $w$
($w\, \epsilon \, [0.1888,\, 1.5482]$) before the  pole.

In figure 3 we restrict the available sequence to higher lattice sizes,
 $h=1/12,\, 1/16,\, 1/20,\, 1/24$ and $1/32$.
 In that case the extrapolation with $0.1\%$ precision is also 
compatible with a still large range of values for $w$:
$w \, \epsilon$ \, $[0.2816,\, 1.1969]$,
before the pole at $w \sim 1.2078$.
 In figure 4, we restrict further the sequence to
 $h=1/16,\, 1/20,\, 1/24$ and $1/32$, and we obtain 
$w \, \epsilon \, [0.411,\, 1.3220]$, and a pole at
$w \sim 1.3871$.

Therefore,  as we restrict our sequences to higher
lattice sizes, the effects of the correction to scaling term
become more pronounced, leading to a smaller range for
acceptable values of $w$. 
 This effect has a stronger counterpart in the criterium of minimum
error: the acceptable range for $w$ is actually narrower than stated
above, mainly for figures 3 and 4. 

We are finally now at a point where we can use the BST algorithm to
analyse our 3D Ising model data of table 2.
  Since we have to discard smaller lattices in order to accomplish 
with the  expected validity conditions to derive  
equations \,(\ref{eq:cross}) and \,(\ref{eq:nu}),
%%  In this way
 we are left with the sequence 
in the third column in table 2.
 Figure 5 presents our  so obtained BST estimates for the critical 
exponent $\nu$ from a sequence for lengths 
 $h=1/6,\,1/8,\, 1/12,\,$ and $1/16$.
  Since our sequences for $1/\nu_{L,2L}$
are restricted to lattice sizes from $L=6$ up to $16$ we take
the four parameter fit  estimate $\nu=0.6280$ and its statistical
error $0.0015$ as our 
input condition to find out $w$, as exemplified for the 2D Ising model.
  This statistical error leads to a range 
in $w$ given by $w= 0.745(74)$, marking the 
 region of minimum systematic error.

Finally, we remark that this approach 
yields a value for $w$ which is in full agreement 
with recent results from various other and 
different approaches: from scaling relations of
observables related to the magnetization \cite{parisi} it was estimated 
$w=0.87(9)$, from perturbative expansion at fixed $D=3$ dimension 
followed $w=0.799(11)$, and 
$w=0.814(18)$ was estimated from $\varepsilon$-expansion \cite{zinn}.

%%%%%%%%%%%%%%%%%%%%%%%%%%%%%%%%%%%%%%%%%%%%%%%%%%%%%%%%%%%%%%%%%

\section{Conclusions}
\noindent

  In summary, we have described a new calculation of the correction to
scaling exponent $w$ based on Monte Carlo multicanonical simulations and 
finite size scaling  theory for the partition function zeros. 
  A four-parameter
fitting was done in order to find out the correlation length critical 
exponent $\nu$. Next, the result for $\nu$ including the statistical error 
was used to obtain the acceptable values for the
renormalization exponent $y_3=-w$ by means of the
BST algorithm. 
 This algorithm helped us to overcome the difficulties in the straight
application of the multiparameter fit (\ref{eq:cross})
to few data points and rather large lattice sizes.

  Our results  for $\nu$ and $w$ are in good 
agreement with recently obtained estimates by Ballesteros et al. \cite{parisi} 
as well as perturbative expansion calculations 
by Guida and Zinn-Justin \cite{zinn}. 

   It is tempting to assume that an
accurate value for $w$ could be pursued by increasing the significant
precision of the complex partition function zeros of the  3D Ising model.
This would  account for a more precise calculation of $\nu_{L,L'}$, 
 by evaluating crossings between lattice sizes $L$ and $L'$.  
  However as we have seen from figures 2 to 4, where we used {\it exact}
 values for $\nu_{L,L'}$, high precision in $\nu$ does not
necessarily lead to a smaller range in $w$. 
  Hence, as an overall conclusion, we note that the large range of $w$ 
is not just a matter of a lack of statistical precision but demonstrates 
that it is necessary to go to much larger lattice sizes.
In particular,  for 2D Ising model even $L=64$ 
seems  to be not large enough.

%%%%%%%%%%%%%%%%%%%%%%%%%%%%%%%%%%%%%%%%%%%%%%%%%%%%%%%%%%%%%%%%%
\vspace{0.4cm}
{\bf Acknowledgements}
\vspace{0.4cm}

  U. Hansmann gratefully acknowledges support by 
the National Science Foundation (CHE-9981874), and 
a generous travel grant by FAPESP (Brazil) which allowed him to
visit  the Ribeir\~ao Preto Campus
of the Universidade de  S\~ao Paulo, Departamento de F{\'{\i}}sica e 
Matem\'atica. N.A.~Alves and J.R. Drugowich de Felicio acknowledge 
support by the brazilian agencies FAPESP, CAPES and CNPq.

%\noindent
%\begin{references}

{
\newpage
\noindent
{\Large Table Captions:} \\

\noindent
{\bf Table 1:}  First partition function zeros 
          for the 3D Ising model on a cubic lattice. \\

\noindent
{\bf Table 2:} Sequence of estimates for the critical exponent 
         $\nu_{L,2L}$  from pairs of lattices  $(L,2L)$. 
         The third column is obtained by replacing
         $|u_1^0(2L)-u_c|\,/ \,|u_1^0(L)-u_c|$  
         by ${\rm Im}\,u_1^0(2L)\,/ \, {\rm Im}\,u_1^0(L)$ 
          in equation \,(\ref{eq:nu}). \\

%%%%%%% begin tables %%%%%%%%%%%%%%%%%%%%%%%%%%%%%%%%%%%%%%%%%%%
\newpage

% TABLE 1
\begin{table}[h]
\renewcommand{\tablename}{Table}
\caption{\baselineskip=0.8cm First partition function zeros 
for the 3D Ising model on a cubic lattice.}
\begin{center}
\vspace{0.2cm}
\begin{tabular}{l c c c c c c}
\hline
\\[-0.3cm]
~$L$ & NRUN & NSWEEP &~~~Re$(u_1^0)$ &~~~Im$(u_1^0)$  &~Re\,$(\beta_1^0) $ 
&~Im\,$(\beta_1^0)$ \\
\\[-0.4cm]
\hline
\\[-0.3cm]
~6 & 2048 & 100\,000& ~~$0.397\,586(18)$ & ~$0.045\,435(17)$ & $0.228\,964(11)$ & 
~$0.028\,445(11)$ \\
~8 & 1024 & 600\,000& ~~$0.402\,723(11)$ & ~$0.028\,596(10)$ & $0.226\,748(07)$ & 
~$0.017\,722(06)$ \\
12 &  512 & 900\,000& ~~$0.407\,018(12)$ & ~$0.014\,925(12)$ & $0.224\,557(07)$ & 
~$0.009\,163(08)$ \\
16 & 512  &1\,200\,000& ~~$0.408\,814(11)$ & ~$0.009\,422(11)$ & $0.223\,557(07)$ & 
~$0.005\,761(07)$ \\
24 & 256  &1\,800\,000& ~~$0.410\,341(14)$ & ~$0.004\,935(12)$ & $0.222\,674(09)$ & 
~$0.003\,006(07)$ \\
32 & 256  & 1\,800\,000&~~$0.410\,991(15)$ & ~$0.003\,124(14)$ & $0.222\,289(09)$ & 
~$0.001\,900(08)$ \\
\\[-0.3cm]
\hline\\
\end{tabular}
\end{center}
\end{table}

\clearpage
% TABLE 2
%%\begin{table}[htbp]
\begin{table}[th]
\renewcommand{\tablename}{Table}
\caption{\baselineskip=0.8cm Sequence of estimates for the critical 
exponent $\nu_{L,2L}$  from pairs of
lattices  $(L,2L)$. The third column is obtained by replacing
         $|u_1^0(2L)-u_c|\,/ \,|u_1^0(L)-u_c|$  
      by ${\rm Im}\,u_1^0(2L)\,/ \, {\rm Im}\,u_1^0(L)$ 
          in equation \,(\ref{eq:nu}).}
\vspace{0.2cm}
\begin{tabular}{c c l}\\
\hline
\\[-0.3cm]
~~~~~$L$  & ~~~$\nu_{L,2L}$ & ~~~$\nu_{L,2L}$  \\ 
\\[-0.35cm]
\hline
\\[-0.3cm]
~~~~~3    & $0.60713(10)$  & $0.60916(11)$  \\
~~~~~4    & $0.61986(13)$  & $0.61831(14)$  \\
~~~~~5    & $0.62451(31)$  & $0.62181(33)$  \\ 
~~~~~6    & $0.62577(44)$  & $0.62272(46)$  \\
~~~~~8    & $0.62719(64)$  & $0.62429(67)$  \\
~~~~12    & $0.6278(14)$   & $0.6263(14)$   \\   
~~~~16    & $0.6270(25)$   & $0.6279(26)$   \\    
\\[-0.3cm]
\hline\\
\end{tabular}
\end{table}
\vfill
%%%%%%% end tables %%%%%%%%%%%%%%%%%%%%%%%%%%%%%%%%%%%%%%%%%%%
}

\clearpage

\vspace{1.2cm}
\noindent
{\Large Figure Captions:} \\

\noindent
{\bf Figure 1:} Four 
          parameter fit for $|u_1^0(L)-u_c|$ in function of $L^{-1/\nu}$
          in the range $L=3-32$.
          The least square method gives $\nu=0.62853(35)$ with $Q=0.89$. \\

\noindent
{\bf Figure 2:} BST estimates of the critical exponent $\nu$
          (full lines) and systematic error (dashed lines),
          as a function of the free parameter $w$, for 2D Ising model.
          The extrapolation is obtained from the finite size sequence 
          with lattice sizes from  $L=4$ up to $L=32$. 
          The right side scale refers to the systematic error. \\

\noindent
{\bf Figure 3:} As in figure 2, BST estimates of the critical exponent $\nu$
          (full lines) and systematic error (dashed lines)
          for 2D Ising model, with lattice sizes from $L=12$ up to $32$. 
          The right side scale refers to the systematic error. \\

\noindent
{\bf Figure 4:} As in figure 2, BST estimates of the critical exponent $\nu$
          (full lines) and systematic error (dashed lines)
          for 2D Ising model, with lattice sizes from
          $L=16$ up to $32$. 
          The right side scale refers to the systematic error. \\

\noindent
{\bf Figure 5:} BST estimates of the critical exponent $\nu$
          (full lines) and  systematic error (dashed lines),
          as a function of the free parameter $w$ for 
          3D Ising model. The sequence is obtained with lattice 
          sizes $L=6,8,12$ and $16$. The right side scale refers to 
          the systematic error. \\

\newpage
%%\cleardoublepage

%FIGURE 1
\begin{figure}[t]
%\begin{figure}[!ht]
\begin{center}
\begin{minipage}[t]{0.95\textwidth}
\centering
\includegraphics[angle=-90,width=0.80\textwidth]{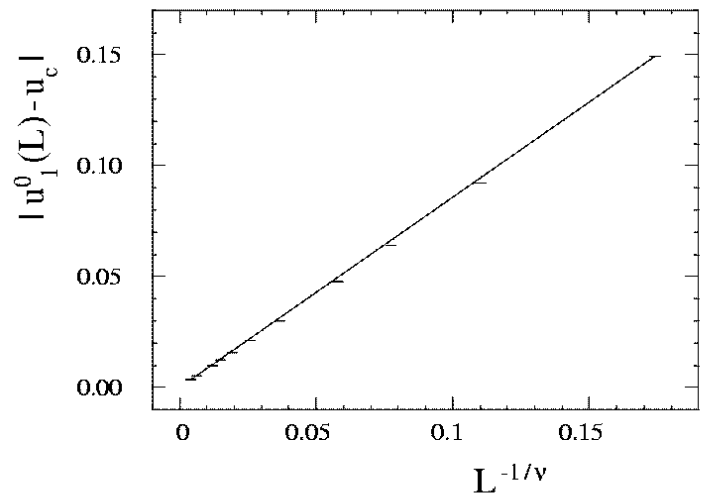}
\renewcommand{\figurename}{Figure}
\caption{Four parameter fit for $|u_1^0(L)-u_c|$ in function of $L^{-1/\nu}$
in the range $L=3-32$.
The least square method gives $\nu=0.62853(35)$ with $Q=0.89$.} 
\label{fig1}
\end{minipage}
\end{center}
\end{figure}

%\newpage
%\cleardoublepage

%FIGURE 2
\begin{figure}[!ht]
%\begin{figure}[t]
\begin{center}
\begin{minipage}[t]{0.95\textwidth}
\centering
\includegraphics[angle=-90,width=0.8\textwidth]{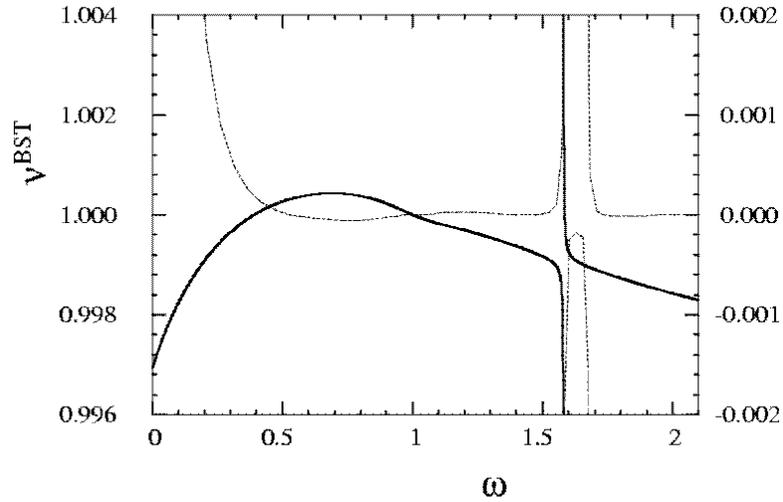}
\renewcommand{\figurename}{Figure}
\caption{BST estimates of the critical exponent $\nu$
 (full lines) and systematic error (dashed lines),
 as a function of the free parameter $w$, for 2D Ising model.
 The extrapolation is obtained from the finite size sequence with
 lattice sizes from  $L=4$ up to $L=32$. 
 The right side scale refers to the systematic error.}
\label{fig2}
\end{minipage}
\end{center}
\end{figure}

%\newpage
%%%\cleardoublepage

%FIGURE 3 
\begin{figure}[t]
%\begin{figure}[!ht]
\begin{center}
\begin{minipage}[t]{0.95\textwidth}
\centering
\includegraphics[angle=-90,width=0.80\textwidth]{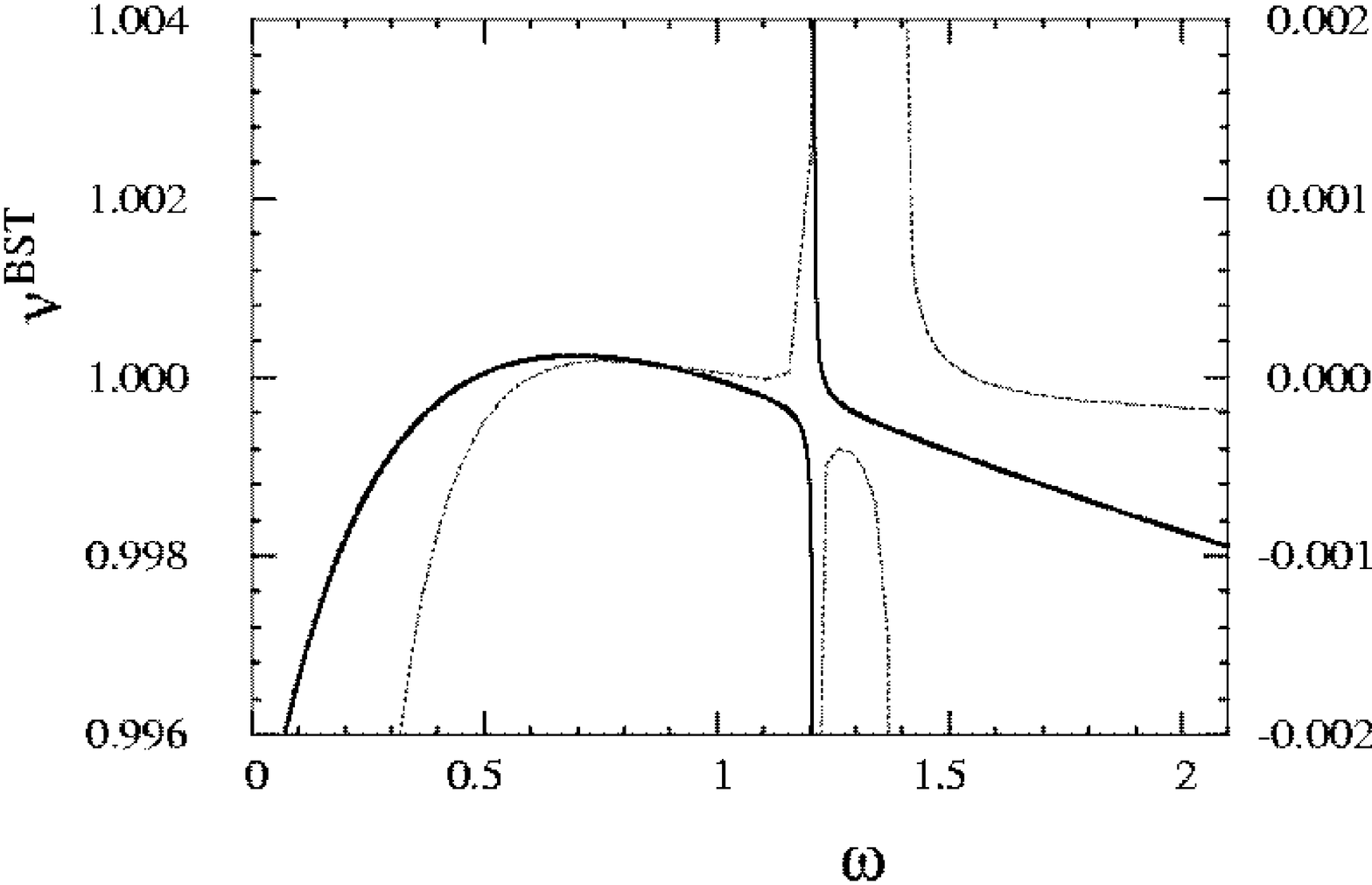}
\renewcommand{\figurename}{Figure}
\caption{As in figure 2, BST estimates of the critical exponent $\nu$
 (full lines) and systematic error (dashed lines)
 for 2D Ising model, with lattice sizes from
$L=12$ up to $32$. The right side scale refers to the systematic error.}
\label{fig3}
\end{minipage}
\end{center}
\end{figure}

%\newpage
%\cleardoublepage

%FIGURE 4
\begin{figure}[!ht]
%\begin{figure}[t]
\begin{center}
\begin{minipage}[t]{0.95\textwidth}
\centering
\includegraphics[angle=-90,width=0.80\textwidth]{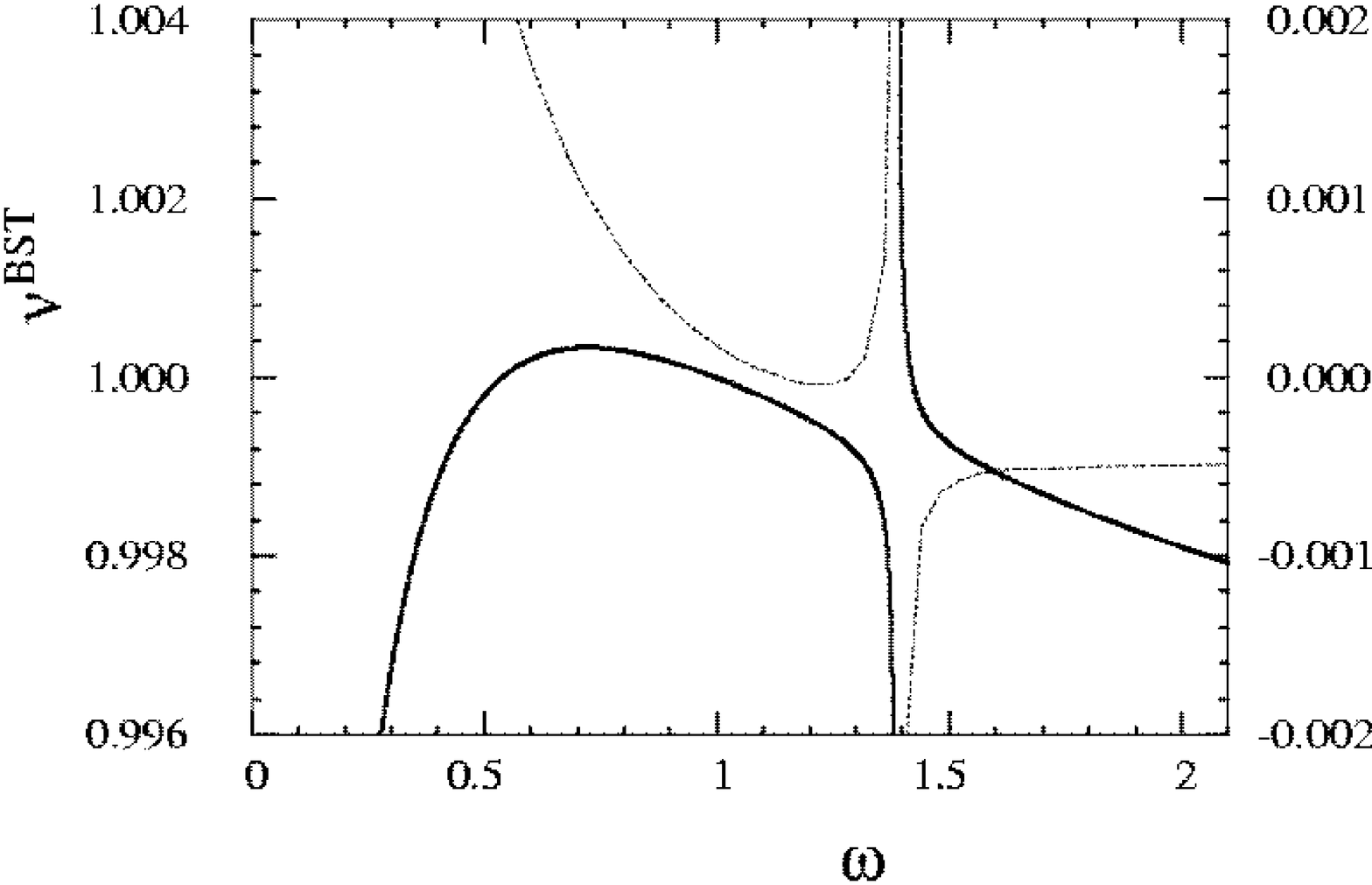}
\renewcommand{\figurename}{Figure}
\caption{As in figure 2, BST estimates of the critical exponent $\nu$
(full lines) and systematic error (dashed lines)
for 2D Ising model, with lattice sizes from
$L=16$ up to $32$. The right side scale refers to the systematic error.}
\label{fig4}
\end{minipage}
\end{center}
\end{figure}

%\newpage
%%%\cleardoublepage

%FIGURE 5
\begin{figure}[!ht]
\begin{center}
\begin{minipage}[t]{0.95\textwidth}
\centering
\includegraphics[angle=-90,width=0.80\textwidth]{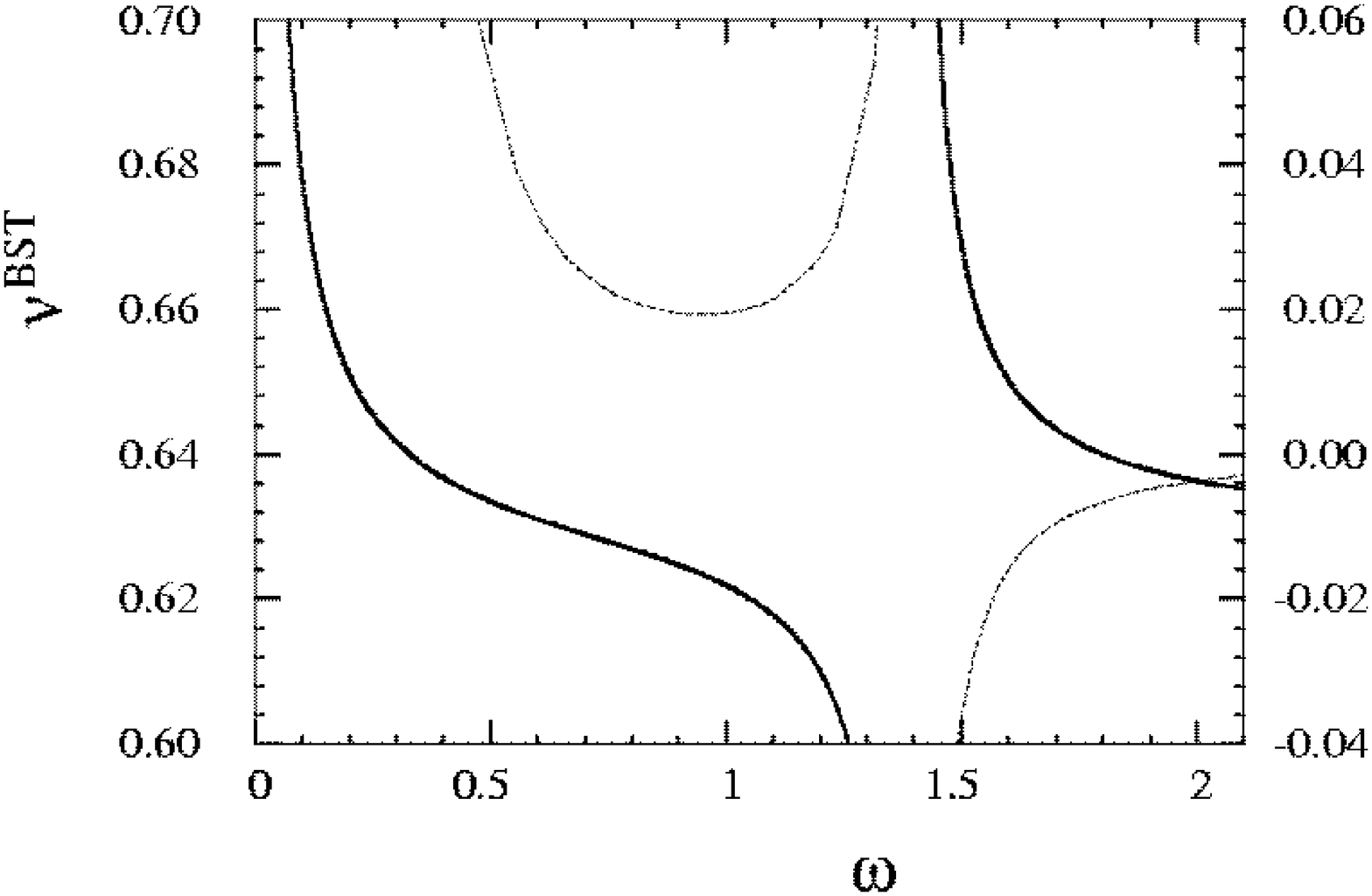}
\renewcommand{\figurename}{Figure}
\caption{BST estimates of the critical exponent $\nu$
 (full lines) and  systematic error (dashed lines),
 as a function of the free parameter $w$ for 
 3D Ising model. The sequence is obtained with lattice sizes $L=6,8,12$ and
 $16$. The right side scale refers to the systematic error.}
\label{fig5}
\end{minipage}
\end{center}
\end{figure}

\begin{thebibliography}{99}

\bibitem{gupta1} Baillie C F, Gupta R, Hawick K A and Pawley G S
                 1992 {\it Phys. Rev.} B {\bf 45} 10438

\bibitem{blote2} Bl\"ote H W J, Luijten E and Heringa J R
                 1995 {\it J. Phys. A: Math. Gen.} {\bf 28} 6289

\bibitem{blote1} Talapov A L and Bl\"ote H W J
                 1996 {\it J. Phys. A: Math. Gen.} {\bf 29} 5727

\bibitem{gupta2} Gupta R and Tamayo P
                 1996 {\it Int. J. Mod. Phys.} C {\bf 7} 305

\bibitem{hasenbusch3} Caselle M and Hasenbusch M
                 1997 {\it J. Phys. A: Math. Gen.} {\bf 30} 4963

\bibitem{hasenbusch2} Hasenbusch M and Pinn K
                 1998 {\it J. Phys. A: Math. Gen.} {\bf 31} 6157

\bibitem{zinn} Guida R and Zinn-Justin J
               1998 {\it J. Phys. A: Math. Gen.} {\bf 31} 8103

\bibitem{parisi}  Ballesteros H G, Fern\'andez L A, 
                  Mart{\'{\i}}n-Mayor V,
                Sudupe A M, Parisi G and Ruiz-Lorenzo J J
                1999 {\it J. Phys. A: Math. Gen.} {\bf 32} 1

\bibitem{hasenbusch1} Hasenbusch M, Pinn K and Vinti S
                   1999 {\it Phys. Rev.} B {\bf 59} 11471

\bibitem{campos}  Campostrini M, Pelissetto A, Rossi P and
                  Vicari E
                  1999 {\it Phys. Rev.} E {\bf 60} 3526

\bibitem{itzykson} Itzykson C, Pearson R B and Zuber J B  
                   1983 {\it Nucl. Phys.} B {\bf 220} [FS8] 415

\bibitem{r13}  Marinari E
               1984 {\it Nucl. Phys.} B {\bf 235} [FS11] 123

\bibitem{ah99b} Alves N A and Hansmann U H E 
               2000 {\it Phys. Rev. Lett.}  {\bf 84} 1836

\bibitem{FS}  Ferrenberg A M and Swendsen R H 
              1988 {\it Phys. Rev. Lett.} {\bf  61} 2635 \\
              1989 {\it Phys. Rev. Lett.} {\bf 63}
              1658(E) and references given in the erratum

\bibitem{salsburg} Salsburg Z W, Jackson J D, Fickett W and Wood W W
             1959 {\it J. Chem. Phys.} {\bf 30} 65 \\
             Chesnut D A and Salsburg Z W
             1963 {\it ibid.} {\bf 38}  2861 \\ 
             McDonald I R and Singer K
             1967 {\it Discuss. Faraday Soc.} {\bf 43} 40 \\
             J. P. Valleau and D. N. Card 
             1972 {\it J. Chem. Phys.} {\bf 57} 5457

\bibitem{swendsen} Ferrenberg A M and Swendsen R H 
             1989 {\it Phys. Rev. Lett.} {\bf  63} 1195

\bibitem{alves1}  Alves N A, Berg B A and Villanova R 
              1990 {\it Phys. Rev.} B {\bf 41} 383

\bibitem{Mu} Berg B A and Neuhaus T
             1991 {\it Phys. Lett.} B {\bf 267} 249

\bibitem{BST}  Bulirsch R and Stoer J 
             1964 {\it Numer. Math.} {\bf 6} 413

\bibitem{s5}  Henkel M and Sch\"utz G 
             1988 {\it J. Phys.} A {\bf 21} 2617

\bibitem{alves2}  Alves N A, Drugowich de Felicio J R  and 
               Hansmann U H E
               1997 {\it Int. J. Mod. Phys.} C {\bf 8} 1063

\bibitem{r14}  Bhanot G, Salvador R, Black S, Carter P and 
               Toral R 
               1987 {\it Phys. Rev. Lett.}  {\bf 59} 803

\bibitem{s1} Brandt S, 1983 
               {\it Statistical and Computational Methods in Data
               Analysis} (Elsevier/North-Holland, New York) 

\bibitem{r16}  Press W H {\it et al} 
               1989 {\it Numerical Recipes:} 
               {\it The Art of Scientific Computing}
               (Cambridge Univ. Press, London)

\bibitem{r24}  Binder K 1981 {\it Z. Phys.} B {\bf 43} 119

\bibitem{s3}   Nightingale M P 1976 {\it Physica} {\bf 83A} 561

\bibitem{r22}  Br\`ezin E 1982 {\it J. Physique} {\bf 43} 15

\bibitem{r23}  Privman V and Fisher M E 
               1983 {\it J. Phys. A: Math. Gen.} {\bf 16} L295

\bibitem{creswick}  Kim S-Y and Creswick R J 
                    1998 {\it Phys. Rev. Let.} {\bf 81} 2000 \\ 
                  Kim S-Y and Creswick R J
                  1999 {\it ibid.} {\bf 82} 3924 \\ 
                    Monroe J L  1999 {\it ibid.} {\bf 82} 3923

\end{thebibliography}
\end{document}